\documentclass[conference]{IEEEtran}
\IEEEoverridecommandlockouts
\usepackage{cite}
\usepackage{amsmath,amssymb,amsfonts}
\usepackage{algorithm}
\usepackage{algorithmic}
\usepackage{graphicx}
\usepackage[caption=false,font=footnotesize]{subfig}
\usepackage{textcomp}
\usepackage[utf8]{inputenc}
\usepackage{xcolor}
\usepackage{tabularx}

\usepackage{graphicx}
\usepackage{fancyhdr}
\usepackage[caption=false,font=footnotesize]{subfig}
\graphicspath{{figures/}}

\def\BibTeX{{\rm B\kern-.05em{\sc i\kern-.025em b}\kern-.08em
    T\kern-.1667em\lower.7ex\hbox{E}\kern-.125emX}}

\begin{document}


\title{Trajectory Sensitivity-Based Dynamic Assessment of an AI Data Center Under Contingencies\\
}

\author{\IEEEauthorblockN{Arnav Bagga, Saskia A. Putri, Abhishek Banerjee, Xiaofan Wu}
\IEEEauthorblockA{\textit{Siemens - Research \& Predevelopment} \\
Princeton, New Jersey-08540, USA \\
abhishek.banerjee@siemens.com, arnav.bagga@siemens.com}
\textcolor{blue}{This work has been accepted for publication in the 58th NAPS conference, 2026 in US.} \\
\textcolor{blue}{Copyright may be transferred without notice, after which this version may no longer be accessible.}} 
\maketitle

\begin{abstract}
Artificial Intelligence (AI) data centers are emerging as converter-dominated, parameter-rich electrical systems whose dynamic response depends on fast workload-driven demand variations, tightly regulated data-hall converters, and coordinated on-site sources. Assessing such systems by repeatedly solving nonlinear time-domain models for every outage, load realization, and parameter perturbation is computationally expensive and provides limited insight into response sensitivity. This paper presents a trajectory sensitivity (TS)-based dynamic assessment framework for a self-contained AI data center supplied by a 2 kV DC microgrid with a grid-tied interface. The model includes time-varying data-hall load profiles, a voltage source converter (VSC), battery energy storage systems (BESS) units, supercapacitors, and synchronous-generation-side sources. For each disturbance, firstly, TS are computed along the nonlinear disturbance trajectory, capturing the local effect of source trips, data-hall outages, converter controls, and load ramps. Later, the computed TS data is used to estimate system trajectories under simultaneous parameter perturbations. Validation against full nonlinear ODE simulations with $\pm 15\%$ perturbations shows close agreement. In addition, the TS-based approach reduces  computational time of dynamic-response evaluations from seconds to milliseconds, demonstrating its potential as a scalable screening tool for AI data center dynamics.
\end{abstract}

\begin{IEEEkeywords}
AI data centers, trajectory sensitivity analysis, dynamic stability, DC microgrids, converter-interfaced loads.
\end{IEEEkeywords}

\section{Introduction}

The rapid growth of hyperscalers and AI-driven data centers is introducing a new class of large, fast-varying loads to modern power systems. Recent estimates indicate that data centers account for approximately 1.5\% of global electricity consumption, with demand expected to increase as AI training and inference workloads continue to expand~\cite{IEA2025}. Unlike conventional aggregate loads, AI data centers are dominated by tightly regulated power-electronic interfaces, hierarchical controls, energy-storage systems, backup sources, and rapidly changing computational demand. These features can produce complex dynamic interactions during voltage sags, source outages, converter saturation, and abrupt load changes~\cite{Sun2022,NERC2025}. Industry reports have further emphasized that emerging large loads can exhibit fast and less predictable power variations, creating new challenges for dynamic security assessment and planning~\cite{NERC2025}.

Recent studies have investigated data center load modeling~\cite{jimenez2025data}, power-system resonance analysis~\cite{zhao2026dynamic}, and voltage and frequency ride-through enhancement~\cite{Wang2026dynamic}. These works show that data center dynamics can differ substantially from conventional static or composite load assumptions, particularly when UPS systems, power converters, cooling demand, and workload-driven ramps are represented. However, dynamic assessment of detailed AI data center models through repeated nonlinear time-domain simulations (TDS) can become computationally burdensome. Each candidate outage, data-hall trip, load-ramp condition, controller setting, or uncertain parameter may require a separate dynamic simulation and post-processing step. This motivates a practical question: how can the post-disturbance response of a parameter-rich AI data center be assessed without repeatedly solving the full nonlinear model for every perturbation.

Trajectory sensitivity (TS) analysis provides a natural framework for this purpose. TS quantifies the time-varying change in system trajectories with respect to parameter perturbations. Unlike small-signal analysis, which linearizes a system around an equilibrium point, TS linearizes the nonlinear response along the simulated trajectory itself~\cite{Hiskens2002}. This property is important for AI data centers because the relevant response occurs during disturbance and post-disturbance intervals, where source trips, data-hall disconnections, converter limits, and fast load ramps may all influence system trajectories. TS has been used in power-system dynamic security assessment, disturbance analysis, contingency screening, preventive control, and parameter influence and estimation studies~\cite{pai2003trajectory,nguyen2014trajectory,Ernst2001,akhlaghi2021trajectory,yuan2019preventive,zhu2024optimal}. However, its application to AI data center dynamic assessment remains largely unexplored.

This paper presents a TS-based framework for examining the dynamic response of a self-contained AI data center supplied through a 2 kV DC microgrid. The studied system includes time-varying data-hall load profiles, a grid-tied VSC, BESS units, supercapacitors, and synchronous-generation-side sources. Disturbance-conditioned TS are computed for selected physical and control parameters and then used to estimate perturbed trajectories under simultaneous parameter changes. This enables fast assessment of monitored states such as DC-bus voltage, source currents, data-hall converter states, and grid-interface variables during high-impact disturbances.

The main contributions of this paper are as follows:
\begin{itemize}
    \item This work lays the foundation for using TS as a dynamic assessment tool for AI data centers with detailed converter-interfaced source and load models.
    \item To the authors' knowledge, this is the first application of TS to an AI data center model subject to time-varying data-hall load profiles and ramps, rather than a fixed-demand or single operating-point assumption.
    \item A disturbance-conditioned TS workflow is developed for data-hall outages and source-side trips, and the resulting TS-predicted trajectories are validated against full nonlinear ODE simulations with simultaneous parameter perturbations.
\end{itemize}

The rest of the paper is organized as follows: Section II provides the system description of the AI data center used in this study. Section III discusses the relevant TS theory from the literature, required to perform dynamic assessment of the AI data center. Section IV explains and validates the results obtained using the TS approach. Finally, Section V provides the conclusion and future work in this area.

\section{AI Data Center System Description}
\begin{figure}[htbp!]
    \centering
    \includegraphics[width=1.0\linewidth]{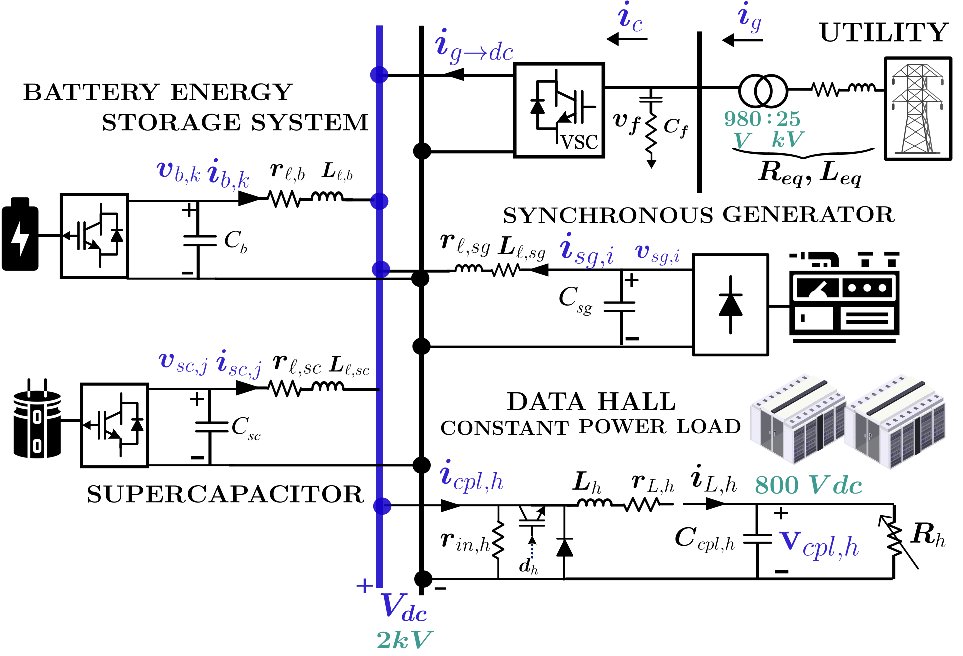}
    \vspace{-15pt}
    \caption{Proposed AI data center power system on a 2 kV DC-bus system}
    \label{fig:layout1}
\end{figure}

\setlength{\abovedisplayskip}{3pt}
\setlength{\belowdisplayskip}{3pt}
The studied grid-connected DC microgrid (MG) for data center power delivery consists of a 2 kV common DC bus supplied by a utility grid via a bidirectional grid-tied voltage-source converter (VSC), and multiple distributed energy resources, including four synchronous generators (SGs) interfaced via rectifier stages, two supercapacitors (SCs), and four battery energy storage systems (BESSs) units connected via bidirectional DC-DC converters, as shown in Fig.~\ref{fig:layout1}. The AI data center load is modeled as two independently varying data hall units, each supplied through tightly regulated buck converters at 800 V, exhibiting constant power load (CPL) behavior. On the utility side, the VSC is connected to the $25 \ kV_{ac}$ grid through an equivalent grid impedance, an AC filter, and a step-down transformer to $980 \ V_{ac}$.

A hierarchical voltage-control strategy is adopted to coordinate the grid-tied VSC with the on-site SG, BESS, and SC units. The grid-tied VSC provides a scheduled base-power contribution and supplementary voltage support. The outer DC-voltage loop generates the grid-to-DC power reference~\cite{marin2018stability}:
\begin{subequations}
\begin{align}
P_{g \rightarrow dc}^{\ast}&= K_{p,g} (v_{ref}-v_{dc})+K_{i,g} \ \xi_{g} + P^\ast_{g \rightarrow dc0}\\
P^\ast_{ac, conv}&= - P_{g \rightarrow dc}^{\ast} \label{eq:Pref}, \ \Rightarrow  i^\ast_{c,d} = \frac{2}{3}\frac{P^\ast_{ac, conv}}{v_{f,d}}, i^\ast_{c,q} = 0 
\end{align}
\end{subequations}
where $\dot \xi_{g}= v_{ref}-v_{dc}$, $v_{dc}$ is the common DC-bus voltage, $v_{ref}$ is the nominal voltage, $P^\ast_{g \rightarrow dc0}$ is the scheduled grid-to-DC power, while the PI terms provide the corrective power needed to regulate the DC-bus voltage. Positive $P_{g \rightarrow dc}^{\ast}$ denotes power imported from the AC grid into the DC bus. The grid-tied VSC is modeled in the synchronous $dq$ reference frame with the upstream grid represented by a Thevenin voltage source behind an equivalent impedance~\cite{marin2018stability}, while the converter-side current is approximated by first-order closed-loop dynamics to track the converter current command $i^\ast_{c,dq}$ defined in~\eqref{eq:Pref}.

For each on-site source $y \in \{b,sc,sg\}$, the voltage error is:
\begin{equation}
    e_{y,m} = v_{ref} + \delta V - \Delta V_{y,m} - v_{y,m}, \label{eq:localVerr}
\end{equation}
where $m$ denotes the unit index, $\delta V$ as common voltage restoration signal generated from a PI controller to minimize bus voltage error with nominal voltage at 2 kV, $v_{y,m}$ represents the local converter terminal voltage, and $\Delta V_{y,m}$ is the virtual droop voltage, assigned based on different dynamic roles to each on-site source, as follows~\cite{hosseinipour2023multifunctional}:
\begin{subequations}
\begin{align}
    \Delta V_{b,k} &= \frac{R_{v,b,k}}{R_{v,b,k}+r_{d,b,k}}\zeta_{b,k}+ \frac{R_{v,b,k}r_{d,b,k}}{R_{v,b,k}+r_{d,b,k}}i_{b,k}, \label{droopbat}\\
    \tau_{b,k} \dot \zeta_{b,k} &= R_{v,b,k} i_{b,k} - \zeta_{b,k},\\
    \Delta V_{sc,j} &= r_{sc,j} i_{sc,j} + v_{csc,j} \ \text{with }  \dot v_{csc,j} = i_{sc,j}/C_{v,sc,j} \label{droopSC}\\
    \Delta V_{sg, i} &= R_{v,sg,i}i_{sg,i}, \label{droopSG}
\end{align}
\end{subequations}
where $\tau_b=(R_{v,b,k}+r_{d,b,k})C_{v,b,k}$. The BESS virtual impedance shapes medium-frequency energy balancing, the SC capacitive droop provides fast transient support, and the SG resistive droop supports base-load sharing.

The controlled source dynamics are represented by averaged terminal-voltage sources connected to the common DC bus through equivalent series $R$--$L$ branches: 
\begin{subequations}
\begin{align}
    L_{\ell,y} \dot i_{y,m} &= v_{y,m} - v_{dc} - r_{\ell,y} i_{y,m} \\
    C_{y,m} \dot v_{y,m} & = i^\ast_{y,m} - i_{y,m}, \\
    i^\ast_{y,m} &= K_{p,y}e_{y,m} + K_{i, y} \vartheta_{y,m}
\end{align}
\end{subequations}
where $\dot \vartheta_{y,m}=e_{y,m}$,  $L_{\ell,y}$ and $r_{\ell,y}$ are the equivalent line inductance and resistance.
\begin{figure}[t!]
    \centering
    \includegraphics[width=1\linewidth]{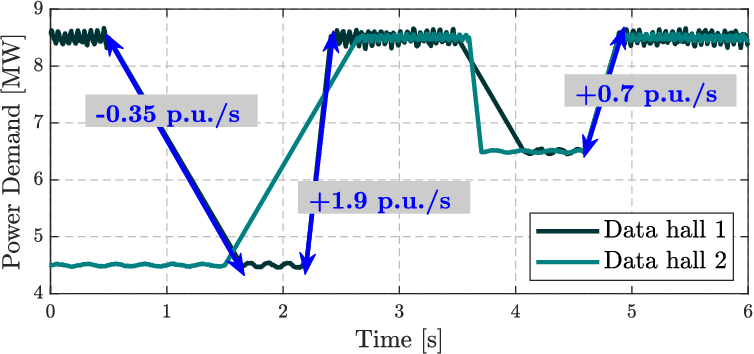}
    \vspace{-.7cm}
    \caption{AI data center load profile, developed from~\cite{NERC2025}}
    \label{fig:loadprof}
\end{figure}

Let $h \in \mathcal{H} = 1,2$ denote the data hall index. The time-varying power demand of the data hall is denoted by $P_h$, used as an exogenous input to the converter-interfaced load model. The profiles in Fig.~\ref{fig:loadprof} are curated to reflect the rapid load-ramping and high noise amplitude characteristics of large-scale data center demand as reported in~\cite{NERC2025}. The converter dynamics are approximated using an averaged buck-converter model that retains the dominant load-side voltage and current dynamics relevant to the DC MG response~\cite{zhao2026dynamic}:
\begin{subequations} \label{eq:loaddyn}
\begin{align}
    \dot i_{L,h} &= \frac{1}{L_{h}} \big(d_h v_{dc} - r_{L,h} i_{L,h} - v_{cpl,h} \big), \\
    \dot v_{cpl,h} &= \frac{1}{C_{cpl,h}} \big(i_{L,h} - \frac{v_{cpl,h}}{R_h} \big),  \ R_h = \frac{(v_{load}^{ref})^2}{P_h},\\
    d_h &= D_{0,h} + K_{p,h}^{cpl} (v_{load}^{ref} - v_{cpl,h}) + K_{i,h}^{cpl} \xi_{cpl,h}\\
    \dot \xi_{cpl,h} &= v_{load}^{ref} - v_{cpl,h} \\
    i_{cpl,h} &= d_h i_{L,h} + \frac{v_{dc}}{r_{in,h}} \label{eq:icpl}
\end{align}
\end{subequations}
where $R_h$ represents the equivalent load resistance, $v_{load}^{ref} = 800 V$ is the regulated load-side voltage reference, $i_{L,h}$ is the converter inductor current, $v_{cpl,h}$ denotes the load-side capacitor voltage, $d_h \in [0,1]$ is the buck-converter duty cycle and $D_{0,h}$ is the nominal duty cycle. The DC-bus input current drawn by the $h$-th data hall is given in \eqref{eq:icpl}, where $r_{in,h}$ denotes a small leakage path at the converter input.
Finally, the common DC bus is modeled using the capacitor current balance, expressed as:
\begin{equation}
C_{eq}\dot{v}_{dc}=\sum_{k=1}^{4} i_{b,k} + \sum_{j=1}^{2} i_{sc,j} +\sum_{i=1}^{4} i_{sg,i}+ i_{g\rightarrow dc}-\sum_{h=1}^{2} i_{cpl,h},\label{eq:dcbus}
\end{equation}
where $C_{eq}$ is the equivalent DC-link capacitance, $i_{b,k}$, $i_{sc,j}$, and $i_{sg,i}$ are the output currents of the $k$-th BESS, $j$-th SC, and $i$-th SG-side source, respectively, and $i_{g\rightarrow dc}$ is the equivalent DC-side current contribution of the grid-tied VSC.


\section{Trajectory Sensitivity-based Method}
In this paper, TS is used to approximate how monitored AI data center trajectories change under selected parameter perturbations during time-varying data-hall loading and contingency events. The objective is not to replace the nonlinear model, but to obtain a trajectory-local linear representation that can rapidly estimate perturbed responses once a disturbed base trajectory has been simulated.

Let $\lambda \in \mathbb{R}^{n_p}$ denote the vector of system's candidate parameters, and $\nu \in \mathcal{F}$ denote a disturbance scenario. For a given disturbance scenario $\nu$, the system trajectory is represented by the flow map
\begin{equation}
    z^{\nu}(t;\lambda)
    =
    \Phi_{\nu}\!\left(t,z_{\mathrm{init}},\lambda\right),
    \label{eq:flow_map}
\end{equation}
where $z(t)\in\mathbb{R}^{n_m}$ contains the monitored states and output variables, $z_{\mathrm{init}}$ is the initial condition, and $\Phi_{\nu}(\cdot)$ denotes the nonlinear time-domain simulation under scenario $\nu$. The nominal trajectory for disturbance scenario $\nu$ is
\begin{equation}
    z_{\mathrm{base}}^{\nu}(t)
    =
    z^{\nu}(t;\lambda^0),
    \label{eq:base_traj}
\end{equation}
where $\lambda^0$ is the nominal parameter vector.

The TS of the monitored states and output variables with respect to the $j$th parameter is defined as
\begin{equation}
    T_j^{\nu}(t)
    =
    \left.
    \frac{\partial z^{\nu}(t;\lambda)}
    {\partial \lambda_j}
    \right|_{\lambda=\lambda^0}.
    \label{eq:ts_definition}
\end{equation}
In this study, TS are computed numerically using a finite-difference procedure. For the $j$th candidate parameter,
\begin{equation}
\begin{split}
    T_j^{\nu}(t_k)
    \approx
    \frac{
    z^{\nu}\!\left(t_k;\lambda^0+
    \Delta\lambda_j^{TS}e_j\right)
    -z^{\nu}\!\left(t_k;\lambda^0\right)}
    {\Delta\lambda_j^{TS}},
\end{split}
\label{eq:finite_difference_ts}
\end{equation}
where $e_j$ is the $j$th unit vector and $t_k$ is a sampled simulation time. The quantity $\Delta \lambda_j^{\mathrm{TS}}$ accounts for the small perturbation done to each of the candidate parameters for TS computation. For this study, we select  2\%, leading to:
\begin{equation}
    \Delta\lambda_j^{TS}
    =
    0.02|\lambda_j^0|.
    \label{eq:ts_step}
\end{equation}
Thus, each sensitivity is obtained from one base simulation and one additional simulation in which only one parameter is perturbed. The same warm-start initial condition is used for all base and perturbed simulations so that the resulting sensitivities reflect parameter influence rather than an initial-condition mismatch.

For each system disturbance scenario, the computed sensitivities are stored as a TS tensor (3-D array),
\begin{equation}
    \mathcal{T}^{\nu}
    \in
    \mathbb{R}^{N_t \times n_m \times n_p},
    \label{eq:ts_tensor}
\end{equation}
where $N_t$ is the number of sampled time instants, $n_m$ is the number of monitored variables, and $n_p$ is the number of candidate parameters. Equivalently, at time $t_k$, the TS matrix can be viewed as
\begin{equation}
    T^{\nu}(t_k)
    =
    \begin{bmatrix}
    T_1^{\nu}(t_k) &
    \cdots &
    T_{n_p}^{\nu}(t_k)
    \end{bmatrix}.
    \label{eq:ts_matrix}
\end{equation}

After $T^{\nu}$ has been computed, a new perturbed trajectory of interest can be estimated without another full nonlinear simulation. For a prescribed parameter perturbation vector $\Delta\lambda$, the first-order TS approximation gives
\begin{equation}
    \widehat{z}^{\nu}(t_k)
    =
    z_{\mathrm{base}}^{\nu}(t_k)
    +
    T^{\nu}(t_k)\Delta\lambda .
    \label{eq:ts_prediction}
\end{equation}
Equivalently,
\begin{equation}
    \widehat{z}^{\nu}(t_k)
    =
    z_{\mathrm{base}}^{\nu}(t_k)
    +
    \sum_{j=1}^{n_p}
    T_j^{\nu}(t_k)\Delta\lambda_j .
    \label{eq:ts_prediction_sum}
\end{equation}
\eqref{eq:ts_prediction} or \eqref{eq:ts_prediction_sum} is the main computational advantage of utilizing TS. Once the disturbance-conditioned TS database is available, several perturbed system responses can be estimated through inexpensive matrix-vector operations. The approximation is physically informative because it is built around the actual pre- and post-disturbance trajectory.

The implementation in this work uses five disturbance scenarios and a total of 35 candidate parameters are selected from the physical, network, and controller parameter sets as listed in Table~\ref{tab:candidate_params_results}. Thus, for each disturbance scenario, one nominal simulation and 35 one-at-a-time perturbed simulations are used to construct $\mathcal{T}^{\nu}$.

Validation is performed by comparing the TS-predicted trajectory \eqref{eq:ts_prediction} with an independent nonlinear simulation. In contrast to the TS-generation stage, the validation case perturbs all selected parameters simultaneously by a larger prescribed amount.
The corresponding nonlinear trajectory is obtained from
\begin{equation}
    z_{ODE}^{\nu}(t_k)
    =
    z^{\nu}\!\left(
    t_k;\lambda^0+\Delta\lambda^{val}
    \right).
    \label{eq:validation_ode}
\end{equation}
A close match between \eqref{eq:ts_prediction} and \eqref{eq:validation_ode} verifies that the computed sensitivities capture the system parameter variations for the selected disturbance. Since the validation perturbation is larger than the perturbation used to compute TS, this comparison also provides a practical check on the useful linear range of the TS approximation. All the discussed steps are summarized in Algorithm 1. 

\begin{algorithm}[t]
\caption{TS-Based Dynamic Assessment of AI Data Center}
\label{alg:ts_assessment}
\footnotesize
\begin{algorithmic}[1]
\REQUIRE Contingency set $\mathcal{F}$, candidate parameter set $\mathcal{P}$,
nominal parameters $\lambda^0$, initial condition $z_{\mathrm{init}}$, time-varying data-hall load profiles
\ENSURE TS tensors $\mathcal{T}^{\nu}$ and validated perturbed trajectories
\FORALL{$\nu \in \mathcal{F}$}
    \STATE Apply disturbance scenario $\nu$
    \STATE Simulate the nonlinear model using $\lambda^0$
    \STATE Store $z_{\mathrm{base}}^{\nu}(t_k)$
    \FORALL{$\lambda_j \in \mathcal{P}$}
        \STATE Set $\Delta\lambda_j^{TS}=0.02|\lambda_j^0|$
        \STATE Perturb only $\lambda_j$
        \STATE Recompute dependent parameters
        \STATE Simulate from the same initial condition
        \STATE Compute $T_j^{\nu}(t_k)$ using \eqref{eq:finite_difference_ts}
    \ENDFOR
    \STATE Assemble the TS tensor $\mathcal{T}^{\nu}$
    \STATE Select simultaneous perturbation vector $\Delta\lambda^{val}$
    \STATE Predict $\widehat{z}_{val}^{\nu}(t_k)$ using \eqref{eq:ts_prediction}
    \STATE Simulate the nonlinear model using $\lambda^0+\Delta\lambda^{val}$
    \STATE Compare TS-estimated and nonlinear trajectories
\ENDFOR
\end{algorithmic}
\end{algorithm}

\section{Numerical Results}
This section presents the TS generation and validation results for the AI data center dynamic model. The simulation horizon was $6$ s, at a $1$ ms time step. Therefore, each base and perturbed run produced a time-discretized dynamic response over the complete pre-, during, and post-disturbance interval. All contingencies were applied at $t=3$ s. The system was coded and simulated in MATLAB using the variable-step stiff ODE solver \textit{ode15s}, with relative tolerance $10^{-5}$ and absolute tolerance $10^{-6}$.

\subsection{Disturbance Scenarios and Recorded States}

The TS tensor $\mathcal{T}^{\nu}$ was generated for five disturbance scenarios, summarized in Table~\ref{tab:faults}. The first case represents a data-hall-side disturbance, while the remaining cases represent source-side outages. For source outages, the tripped source current was forced to decay with a $1$ ms time constant, which emulates fast disconnection while avoiding numerical discontinuities.

\begin{table}[t]
\centering
\caption{Disturbance scenarios used for TS generation.}
\label{tab:faults}
\footnotesize
\begin{tabularx}{\columnwidth}{cX}
\hline
\hline
Case & Disturbance \\
\hline
\hline
F1 & Hall 1 permanent outage; Hall 2 temporary outage (disconnects at 3 s \& reconnects at 4.5 s) \\
F2 & BESS $\#$1 trip \\
F3 & SG $\#$ 1 trip \\
F4 & SC $\#$1 trip \\
F5 & Simultaneous BESS $\#$1 and SG $\#$1 trip \\
\hline \hline
\end{tabularx}
\vspace{-0.3cm}
\end{table}

For each contingency, the recorded state vector included the DC-link voltage, currents and internal states for each sources, CPL converter states, secondary-control integrator, and grid-interface $dq$ states. In total, $51$ dynamic states ($n_m$) were recorded at every simulation time step. 

The candidate parameters used for TS generation are listed in Table~\ref{tab:candidate_params_results}. These parameters were selected to span the main physical and control subsystems of the AI data center, including storage-source dynamics, source controllers, secondary voltage control, grid-interface control, CPL control, and line parameters.

\begin{table}[t]
\centering
\caption{Candidate parameters used for TS generation.}
\label{tab:candidate_params_results}
\footnotesize
\begin{tabular}{llc}
\hline
\hline
Group & Parameters & Count \\
\hline
\hline
DC reference/link & $v_{ref}$, $C_{eq}$ & 2 \\
BESS virtual dynamics \& control & $R_{v,b,1:4}$, $C_{v,b,1:2}$, $K_{p,b}$, $K_{i,b}$  & 10 \\
SC dynamics \& control & $r_{sc}$, $C_{v,sc,1:2}$, $K_{p,sc}$, $K_{i,sc}$ & 5 \\
SG virtual resistance \& control & $R_{v,sg,1:4}$,  $K_{p,sg}$, $K_{i,sg}$ & 7 \\
Secondary control & $K_{p,sec}$, $K_{i,sec}$ & 2 \\
CPL control & $K_{p,1:2}^{cpl}$, $K_{i}^{cpl} $ & 3 \\
Network & $r_{\ell}$, $L_{\ell}$ & 2 \\
\hline
Total & & 35 \\
\hline
\hline
\end{tabular}
\vspace{-0.3cm}
\end{table}

\subsection{Trajectory Sensitivity Generation}

For each disturbance scenario, one nominal simulation was first performed to obtain the base disturbance trajectory. Then, each candidate parameter was perturbed individually by $2\%$ of its nominal value, and a new nonlinear simulation was performed from the same warm-start initial condition. The TS were then computed from the difference between the perturbed and base trajectories, divided by the corresponding parameter perturbation magnitude \eqref{eq:ts_definition}. Thus, for each system disturbance scenario, $35$ ($n_p$) one-at-a-time perturbed simulations were used to construct $\mathcal{T}^{\nu}$; for the present study, $\mathcal{T}^{\nu}\in\mathbb{R}^{6001 \times 51 \times 35}$, with dimensions corresponding to time samples, monitored states, and candidate parameters, respectively.

\begin{figure*}[t]
\centering
\includegraphics[width=0.9\textwidth]{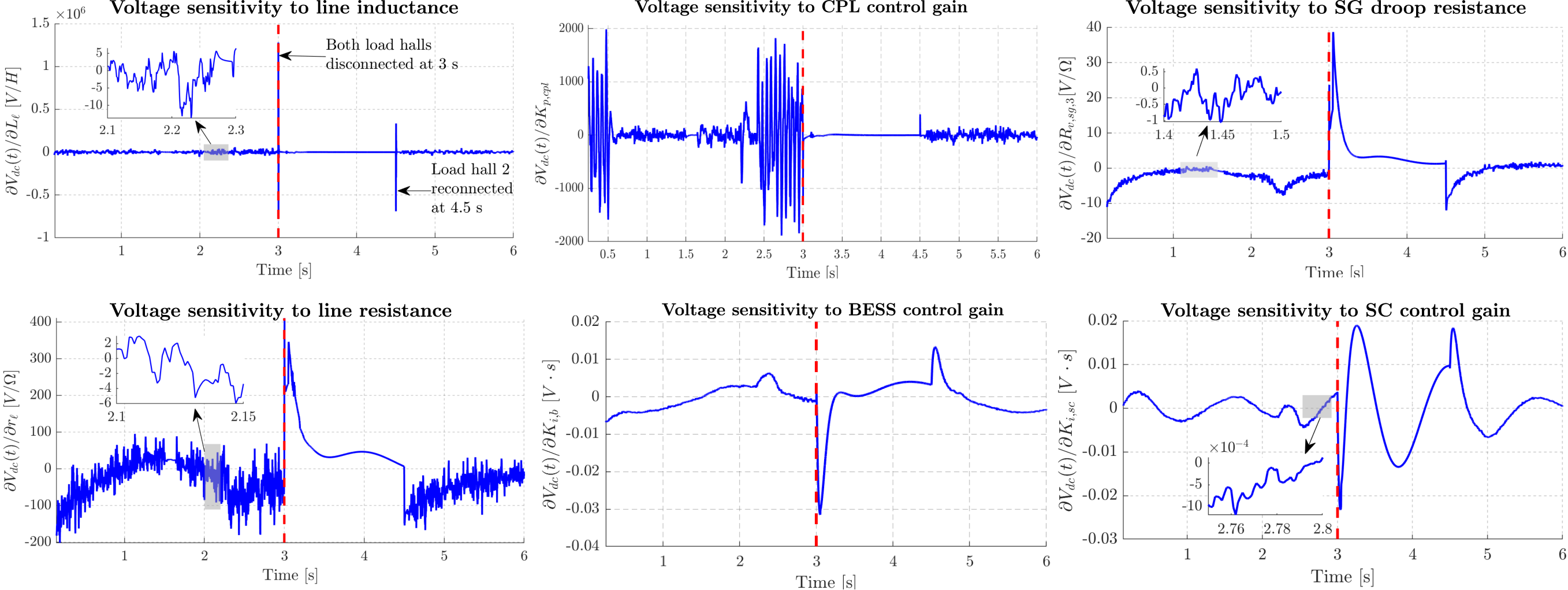}
\caption{TS time-series obtained for the DC-bus voltage subject to different candidate parameters for case F1.}
\label{fig:ts_vdc_params}
\end{figure*}

Representative TS plots are shown in Fig.~\ref{fig:ts_vdc_params}. For clarity, only a few TS profiles of $v_{dc}$ with respect to selected candidate parameters are shown; the complete study computes $51 \times 35 = 1785$ TS profiles per disturbance scenario, or $1785 \times 5 = 8925$ profiles across all five disturbance scenarios. The full set of TS profiles is used later to estimate perturbed system trajectories under simultaneous parameter changes. The example TS plots illustrate the time-varying nature of the computed TS, confirming that TS capture the dynamic effect of both the disturbance and post-disturbance conditions rather than only the pre-disturbance operating condition. It should be noted that in our case, even the pre-disturbance operating condition is subject to consistent load variation.





\begin{figure*}[t]
\centering
\includegraphics[width=0.95\textwidth]{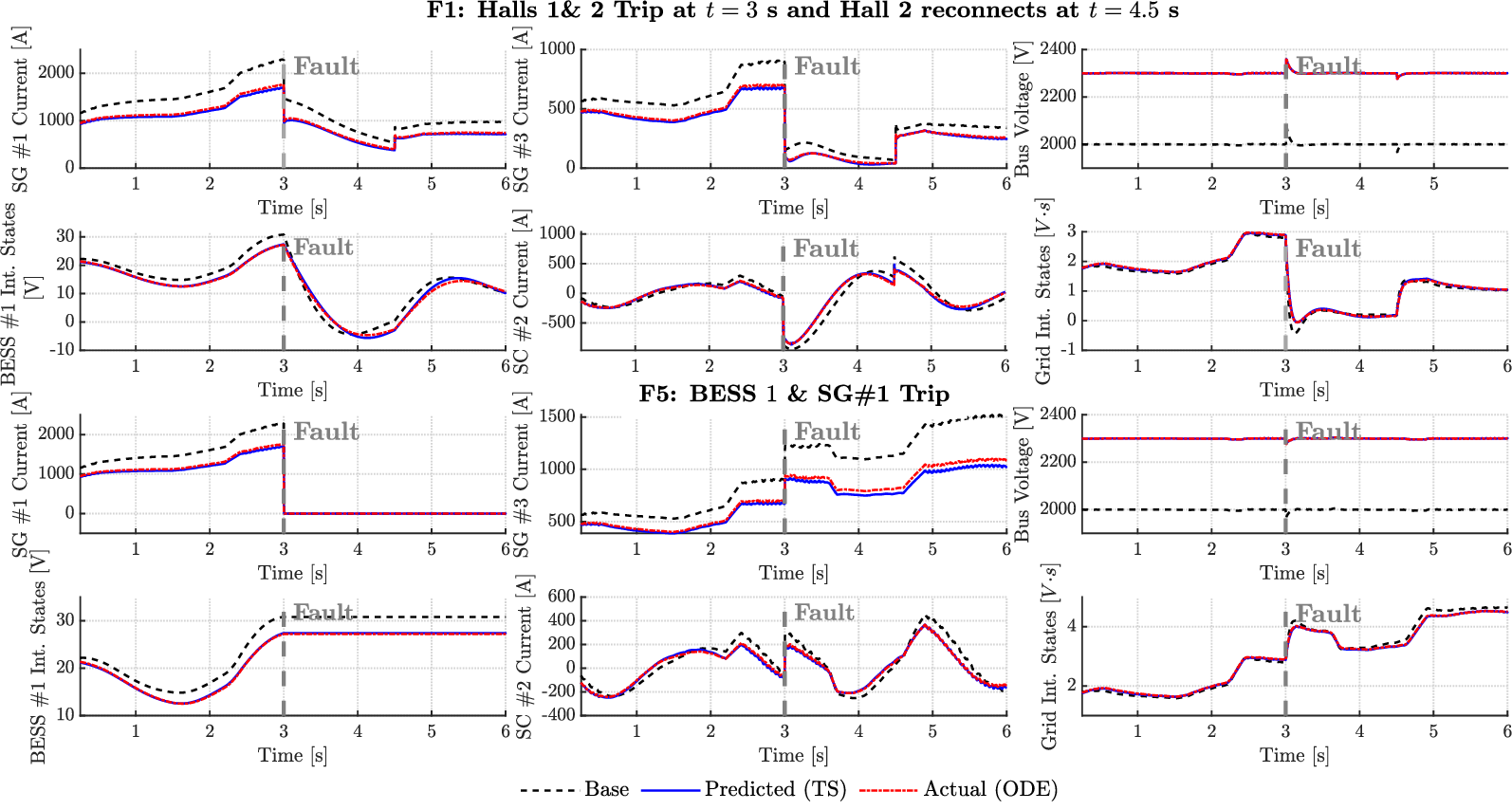}
\caption{Example validation plots for system disturbance scenarios. First two rows: Case F1, where both Hall trips and Hall 2 reconnect at 4.5$s$. Last two rows: Case F5, corresponding to the simultaneous trip of the SG and BESS. All 35 candidate parameters were perturbed by $\pm 15\%$.}
\label{fig:validation_set}
\vspace{-0.5cm}
\end{figure*}

\subsection{Validation With Simultaneous Parameter Perturbations}

After constructing the TS database, validation was performed using parameter perturbations. In contrast to the TS-generation step, where one parameter was perturbed at a time by $2\%$, the validation case perturbed all $35$ candidate parameters together by $\pm 15\%$ according to a prescribed sign vector, for all 5 disturbance scenarios. The TS-predicted trajectory was computed using $\mathcal{T}^{\nu}$ and the perturbation vector \eqref{eq:ts_prediction}. An independent full nonlinear ODE simulation was then performed using the same simultaneously perturbed parameter set \eqref{eq:validation_ode}.

\begin{figure}[htbp!]
\centering
\includegraphics[width=0.8\columnwidth]{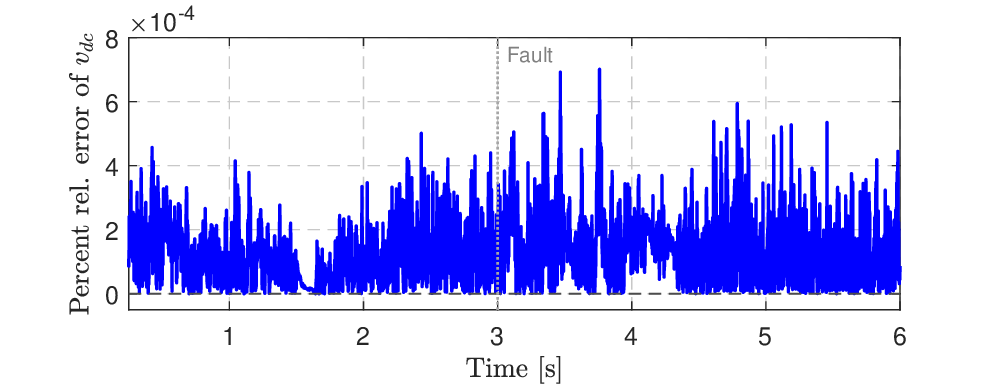}
\vspace{3pt}
\includegraphics[width=0.8\columnwidth]{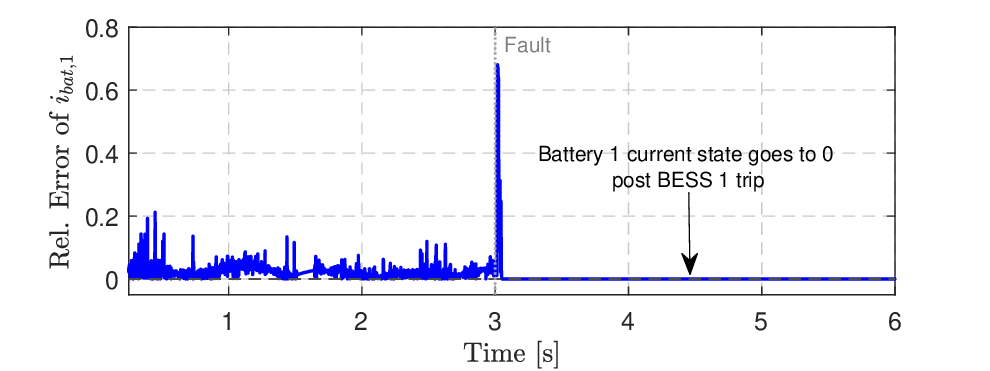}
\vspace{3pt}
\includegraphics[width=0.8\columnwidth]{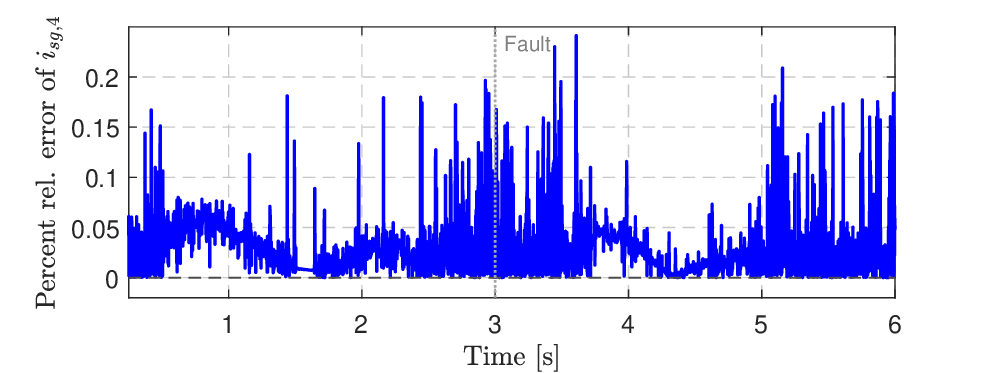}
\vspace{3pt}
\includegraphics[width=0.8\columnwidth]{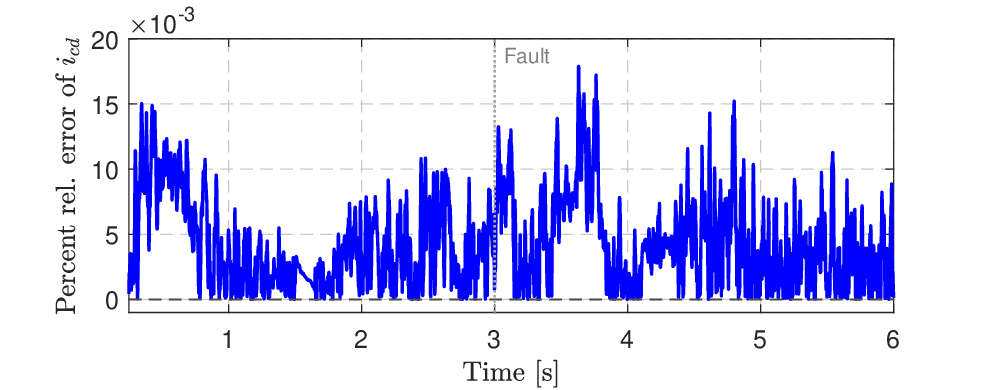}
\vspace{-0.5cm}
\caption{Percentage relative error plots between TS-based state estimates and states obtained from full nonlinear ODE simulation, for F2 or BESS $\# 1$ trip}
\label{fig:error_plots}
\end{figure}

Fig.~\ref{fig:validation_set} shows representative validation plots. In each plot, the base trajectory, TS-predicted trajectory, and full nonlinear ODE trajectory are compared. The close match between \eqref{eq:ts_prediction} and \eqref{eq:validation_ode} across the studied cases indicates that the computed TS capture the dominant dynamic impact of parameter variations under the selected disturbances.
Some examples of corresponding prediction-error traces for the study are shown in Fig.~\ref{fig:error_plots}. The error is evaluated pointwise as the relative (percentage) deviation between the nonlinear ODE trajectory and the TS-based prediction, given by
\begin{equation}
    \frac{\left| z_{ODE}^{\nu}(t_k) - \widehat{z}_{val}^{\nu}(t_k) \right|}{\left| z_{ODE}^{\nu}(t_k) \right|} \times 100\%.
\end{equation}
This metric quantifies, at each time instant, the normalized discrepancy between the actual nonlinear response and its first-order sensitivity-based approximation. The errors remain bounded over pre- and post-disturbance interval, supporting the validity of the TS approximation for fast screening of perturbed dynamic responses.
\subsection{Computational Advantage}
The main practical advantage of the TS approach is that the expensive nonlinear simulations can be complemented by an offline data-driven framework. For the system considered here, a full nonlinear $6$~s dynamic simulation, sampled at $1$~ms resolution, required approximately $10$~s of wall-clock time. 

Once the TS tensor $T^{\nu}$ is available, however, a new perturbed trajectory can be computed using the linearized TS update \eqref{eq:ts_prediction} in as little as $25$~ms. This corresponds to a speed-up factor of approximately $400\times$ relative to the full nonlinear simulation, for this case study.

Although the present case study includes only two varying data-hall units, the computational gains can become significant for planning and screening studies of larger AI data centers. In practical studies, a large number of parameter combinations and operating conditions must be evaluated across multiple disturbance scenarios, making the TS-based approach particularly attractive from a computational standpoint.

\section{Conclusion and Future Work}

This work demonstrates that TS can provide a fast and informative surrogate for repeated nonlinear dynamic simulations of AI data center power systems under time varying extreme load changes. Across the studied data-hall and source-side disturbances, the TS-based trajectory estimates closely tracked the corresponding nonlinear ODE responses when all candidate parameters were perturbed simultaneously. Since a new perturbed response can be recovered from the precomputed TS tensor in milliseconds, the approach is well-suited for rapid post-disturbance screening where many parameter sets, outage combinations, or workload realizations must be evaluated.

Future work will focus on extending TS from trajectory prediction to broader data center dynamic assessment tasks. One important direction is parameter estimation and model calibration for model-heavy data centers, where converter controls, virtual impedances, storage dynamics, and load-side controller parameters may be uncertain or vendor-limited. TS can also be used to identify poorly observable or highly influential parameters, guide sensor placement, and prioritize which subsystem models require higher fidelity. Another direction is adaptive TS database construction for changing workload profiles, allowing sensitivities to be updated only when the operating trajectory changes materially. Finally, applying TS to larger multi-bus data center campuses with AC-side contingencies will help evaluate its usefulness for planning-scale dynamic security studies.


\bibliographystyle{ieeetr}
\bibliography{ref}

\end{document}